\documentclass[apl,twocolumn,floatfix,superscriptaddress,citeautoscript]{revtex4-1}

\usepackage{graphicx}
\usepackage{dcolumn}   
\usepackage{bm}        
\usepackage{amssymb}   
\usepackage{verbatim}
\usepackage{multirow}

\usepackage{amsmath}
\usepackage{amsfonts}

\begin{document}

\title{Microstrip superconducting quantum interference device amplifiers with submicron Josephson junctions: enhanced gain at gigahertz frequencies}

\author{M.P. DeFeo}
\affiliation{Department of Physics, Syracuse University, Syracuse, NY 13244-1130}
\author{P. Bhupathi}
\affiliation{Department of Physics, Syracuse University, Syracuse, NY 13244-1130}
\author{K. Yu}
\affiliation{Department of Physics, Syracuse University, Syracuse, NY 13244-1130}
\author{T.W. Heitmann}
\affiliation{Department of Physics, Syracuse University, Syracuse, NY 13244-1130}
\author{C. Song}
\affiliation{Department of Physics, Syracuse University, Syracuse, NY 13244-1130}
\author{R. McDermott}
\affiliation{Department of Physics, University of Wisconsin, Madison, WI 53706}
\author{B.L.T. Plourde}
\email[]{bplourde@phy.syr.edu}
\affiliation{Department of Physics, Syracuse University, Syracuse, NY 13244-1130}

\date{\today}

\begin{abstract}
We present measurements of an amplifier based on a dc superconducting quantum interference device (SQUID) with submicron Al-AlOx-Al Josephson junctions. 
The small junction size reduces their self-capacitance and allows for the use of relatively large resistive shunts while maintaining nonhysteretic operation. This leads to an enhancement of the SQUID transfer function compared to SQUIDs with micron-scale junctions. 
The device layout is modified from that of a conventional SQUID to allow for coupling signals into the amplifier with a substantial mutual inductance for a relatively short microstrip coil. 
Measurements at 310 mK exhibit gain of 32 dB at 1.55 GHz. 
\end{abstract}

\maketitle

In recent years there have been many advances with amplifiers based on dc superconducting quantum interference devices (SQUIDs) with a resonant stripline input circuit \cite{mueck1998}. 
In these devices the signal is coupled to the SQUID through the $\lambda/2$ microstrip resonance formed by a superconducting spiral input coil above the dielectric layer on top of the superconducting washer that forms the SQUID loop. 
Such amplifiers have exhibited gains in excess of 20 dB in the radiofrequency range \cite{mueck2010} and noise temperatures at 500 MHz within a factor of two of the quantum limit \cite{mueck2001}. 
In addition, microstrip SQUID amplifiers have been demonstrated at frequencies up to 7.4 GHz \cite{mueck2003}. This suggests the possibility of using these devices for measuring the weak signals involved in various quantum information processing schemes with superconducting circuits \cite{clarke2008}, including dispersive readout with circuit quantum electrodynamics \cite{wallraff04} and schemes involving pulsed interactions between qubits and oscillators \cite{serban08}. 
However, the shorter coils required to increase the operating frequency lead to decreased gain: 12 dB at 2.2 GHz and 6 dB at 7.4 GHz \cite{mueck2003}. An alternative configuration with a small-area SQUID coupled in a lumped-element configuration to a quarter-wave resonator was shown to operate as an amplifier in the GHz range \cite{spietz2008}.

\begin{figure}[hb]
\centering
\includegraphics[width=3.35in]{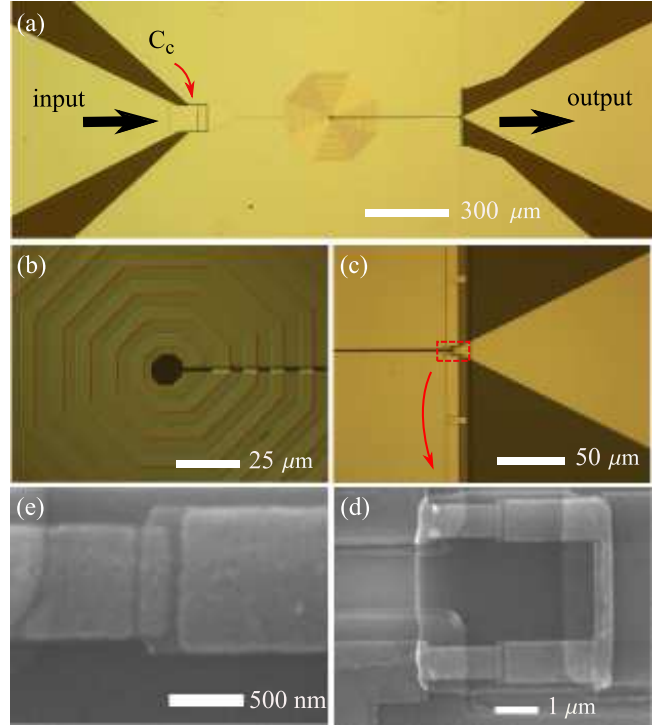}
  \caption{(Color online) (a) Optical micrograph of washer showing coil, input and output ports, and input coupling capacitor. (b) Closeup of input coil. (c) Closeup of junction and shunt region. (d) Scanning electron micrograph of junctions. (e) Closeup of single junction.
\label{fig:setup}}
\end{figure}

The gain $G$ of a microstrip SQUID amplifier is proportional to $M_i^2 V_{\Phi}^2$, where $M_i$ is the mutual inductance between the input coil and the SQUID loop and $V_\Phi \equiv \partial V/\partial \Phi$ is the maximum voltage modulation of the SQUID. 
Pushing the operating frequency $f_0$ higher requires shorter coils, which necessarily reduces $M_i$, although 
this reduction can be mitigated  
somewhat by modifying the SQUID loop and coil layout from the conventional washer design. Nonetheless, $G$ will decrease as $f_0$ is increased unless one can simultaneously compensate by increasing $V_{\Phi}$. 
The peak-to-peak voltage modulation of a SQUID is limited by the $I_0R$ product of each junction, where $I_0$ and $R$ are the junction critical current and shunt resistance, respectively. Nonhysteretic device operation requires a junction damping parameter $\beta_C \equiv (2\pi I_0R/\Phi_0)R C \leq 1$, where $\Phi_0 \equiv h/2 e$, thus placing an upper limit on $R$, where $C$ is the junction self-capacitance. 
For Josephson junctions fabricated with conventional photolithography with an area of a few $\mu{\rm m}^2$, $C$ is typically a few hundred fF. 
The other standard SQUID optimization, $\beta_L \equiv 2 L I_0/\Phi_0 \approx 1$ \cite{tesche77} constrains the product of the SQUID self-inductance $L$ and $I_0$. 
Taking a typical set of dc SQUID parameters, $L=350\,$pH, $I_0=3\,\mu{\rm A}$, $C=200\,$fF, the maximum value of $R$ that maintains nonhysteretic operation is $23\,\Omega$. 
This then results in a maximum flux-to-voltage transfer coefficient $V_{\Phi} \approx R/L =140\,\mu\,$V/$\Phi_0$ \cite{tesche77}. 
To enhance $V_{\Phi}$, one can reduce $L$ somewhat, but it then becomes difficult to avoid loss of gain due to the resulting reduction in $M_i$.

Larger values of $V_\Phi$ can be achieved by increasing $R$, however, $C$ must be reduced in order to avoid hysteretic behavior. 
In this letter, we describe the fabrication of dc SQUIDs with submicron Josephson junctions patterned with electron-beam lithography 
and a device layout tailored to maintain large $M_i$ for short coil lengths. 
We present measurements of one of these SQUIDs operated as a microstrip SQUID amplifier, with gain in excess of 30 dB at $1.55\,$GHz. 

Our SQUID loop consists of a large Al washer on an oxidized Si wafer with a $12\,\mu$m-wide octagonal hole in the center and a $2\,\mu$m slit of length $466\,\mu$m extending to one side. Applying the standard washer-SQUID expressions \cite{ketchen1982} leads to an estimate for the SQUID inductance $L \approx 160\,$pH. 
The Al washer has an outer-width of 6.5 mm at its midpoint and also serves as the ground plane when the SQUID is operated as a microstrip SQUID amplifier, with cutouts allowing for the input and output traces to be coupled in a coplanar-waveguide geometry [Fig. \ref{fig:setup}(a)]. 
The dielectric layer on top of the washer is formed from a $150\,$nm-thick SiO$_2$ film deposited by PECVD.
The Al input coil has a $5\,\mu$m linewidth and follows an octagonal path around the washer hole with a length of $8.3\,$mm and a number 
of turns $n = 16$ [Fig. \ref{fig:setup}(b)]. 
Our present design does not have a connection to the center turn of the coil, thus, a direct dc measurement of $M_i$ is not possible. Nonetheless, for our geometry we can estimate $M_i \approx 1\,$nH \cite{ketchen92}. Between the input pad and the coil, we fabricated an on-chip input coupling capacitor that we estimate to be $C_c \approx 4\,$pF using the same dielectric layer as on the washer to reduce the loading from the $50\,\Omega$ environment on the microstrip resonance \cite{kinion08}.

While the initial four layers of the SQUIDs are patterned photolithographically, the junctions are patterned in a final electron-beam lithography step and are formed with a double-angle shadow-evaporation process \cite{dolan1977}. An in situ Ar ion mill step ensures superconducting contacts between the junction layer and the washer. 
The junctions are 730 nm $\times$ 180 nm [Fig. \ref{fig:setup}(d-e)], from which we estimate the capacitance to be roughly 15 fF. 
The resistive shunts are formed from a 20 nm-thick Pd layer with a low-temperature sheet resistance of 6.1 $\Omega/\square$ 
resulting in $R = 56\,\Omega$. 

\begin{figure}
\centering
\includegraphics[width=3.35in]{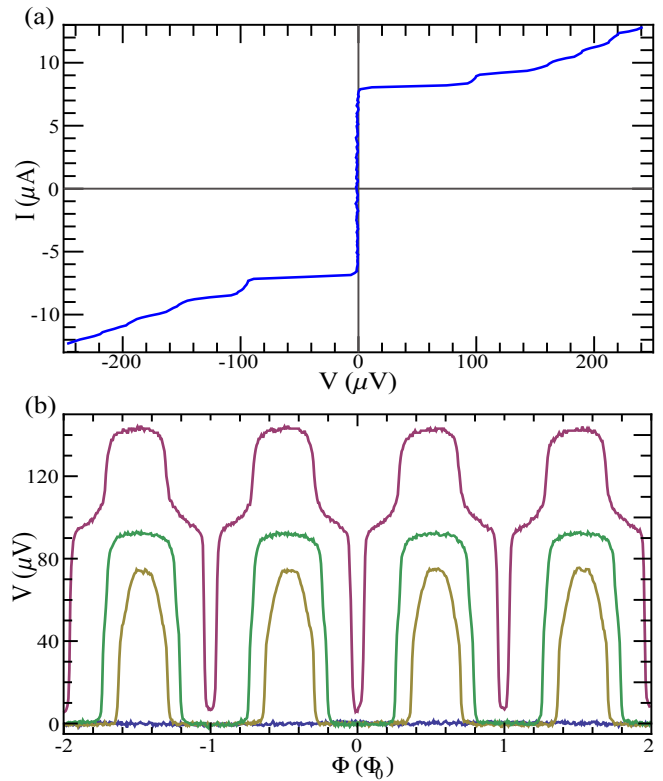}
  \caption{(Color online) (a) Low-frequency current-voltage characteristic at 310 mK for flux bias $\Phi=n \Phi_0$. 
  (b) Flux modulation of voltage across SQUID for different bias current values. 
\label{fig:IVCsVPhis}}
\end{figure}

Prior to measuring the microwave response of the amplifier, we recorded the 
current-voltage characteristics (IVCs) at 310 mK on a separate cooldown of our $^3$He refrigerator [Fig. \ref{fig:IVCsVPhis}(a)]. From the IVCs, we observe $I_0=4.0\,\mu$A, corresponding to $\beta_c \approx 0.6$. Based on a one-parameter fit to the critical current modulation (not shown), we obtain $\beta_L = 0.65$, which, combined with $I_0$ is in reasonable agreement with our earlier estimate for $L$. 
Measurements of the flux modulation of the dc voltage for different bias currents 
[Fig. \ref{fig:IVCsVPhis}(b)] allow us to extract $V_{\Phi}$, with a maximum value of $\sim 3\,$mV/$\Phi_0$. 
We note that during subsequent measurements of gain on this device, $V_{\Phi}$ was likely reduced somewhat due to noise fed back to the SQUID from the microwave post-amplifier.

\begin{figure}[t]
\centering
\includegraphics[width=3.35in]{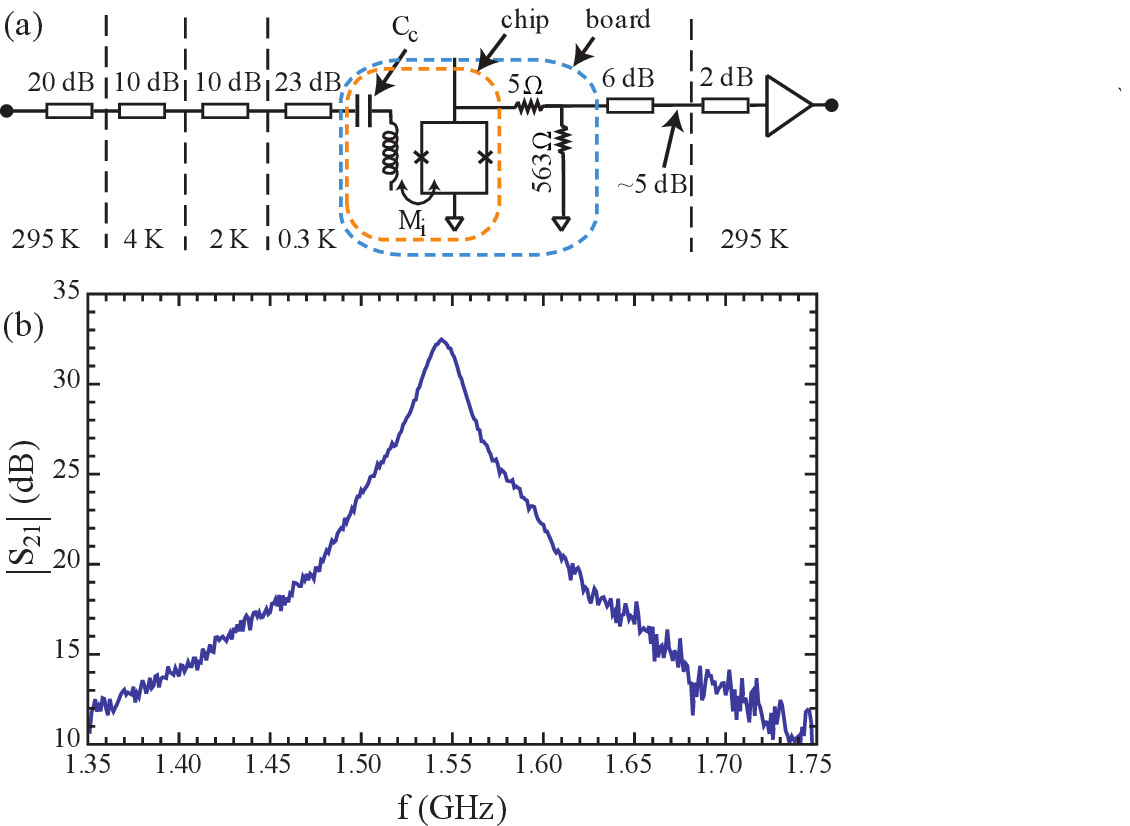}
  \caption{(Color online) (a) Schematic for gain measurement. (b) Measurement of gain, $|S_{21}|(f)$, for optimum current and flux bias. 
  \label{fig:gain-curves}}
\end{figure}

On a subsequent cooldown to 310 mK, we mounted the SQUID on a board with stripline traces attached with multiple short wirebonds to the input and output pads of the SQUID and multiple ground wirebonds to the washer. 
The bias current and flux bias for the SQUID were supplied with batteries and the lines passed through cryogenic Cu-powder filters.  
To shield the SQUID from external magnetic fields, the board was mounted in a closed Al box that was wrapped in Pb foil and a cryogenic $\mu-$metal shield surrounded the vacuum can of the refrigerator. 
The microwave path consisted of multiple stages of attenuation on the input side, including $-23\,$dB anchored to the $^3$He stage, for  attenuating room-temperature noise, followed by a $6\,$dB attenuator on the SQUID output for matching to $50\,\Omega$ [Fig.~\ref{fig:gain-curves}(a)]. For further amplification we used a room-temperature post-amplifier (3 x MiniCircuits ZX60-33) with a gain of 48 dB at 1.55 GHz. 
A $2\,$dB attenuator at the input was necessary to help with matching the ZX60-33 to $50\,\Omega$. 
We calibrated the various cable loss contributions, attenuation, and post-amplifier gain with a short coax piece in place of the SQUID board by measuring the transmission with a network analyzer during a separate cooldown of the refrigerator. Subsequent gain measurements of the SQUID were referred to this baseline.

We measured the gain of the SQUID amplifier using a network analyzer to supply a weak power to the input, $\sim -120$ dBm (Fig. \ref{fig:gain-curves}). For the optimum bias current and flux values, we measured a maximum gain of 32 dB at 1.55 GHz with a bandwidth of 30 MHz. 
Upon tuning to the optimum point, the gain was stable and there was no evidence for self-oscillations that can sometimes be present for microstrip SQUID amplifiers under certain biasing and input conditions \cite{kinion10}. 
A measurement of the noise temperature was not practical with the present configuration due to the substantial contribution from the room-temperature post-amplifier. 
Future measurements with a cryogenic HEMT amplifier should greatly reduce the noise contribution of the post-amplifier. 

One potential concern with increasing the shunt size is the possibility for hot-electron effects leading to an elevated temperature for the shunts \cite{wellstood89, wellstood94}. This can be addressed to some extent with metallic cooling fins added to the shunts \cite{wellstood89}. In addition, the enhanced gain resulting from the larger shunts should at least partially compensate any excess noise due to 
hot-electron effects in the shunts.

In conclusion, we have fabricated microstrip SQUID amplifiers with Al-AlOx-Al submicron junctions. From low frequency measurements, we observed quite large $V_\Phi$, which, combined with large $M_i$ leads to stable operation with gain in excess of 30 dB at 1.55 GHz with a bandwidth of 30 MHz. 
Shortening the coils on future devices with the present design should maintain sufficiently large $M_i$ to allow for substantial gain at frequencies up to at least several GHz. 

This work is supported by the DARPA/MTO QuEST program through a grant from AFOSR. Some of the device fabrication was performed at the Cornell NanoScale Facility, a member of the National Nanotechnology Infrastructure Network, which is supported by the National Science Foundation (Grant ECS-0335765). The authors acknowledge M. Ware for technical assistance.


\begin{thebibliography}{16}
\expandafter\ifx\csname natexlab\endcsname\relax\def\natexlab#1{#1}\fi
\expandafter\ifx\csname bibnamefont\endcsname\relax
  \def\bibnamefont#1{#1}\fi
\expandafter\ifx\csname bibfnamefont\endcsname\relax
  \def\bibfnamefont#1{#1}\fi
\expandafter\ifx\csname citenamefont\endcsname\relax
  \def\citenamefont#1{#1}\fi
\expandafter\ifx\csname url\endcsname\relax
  \def\url#1{\texttt{#1}}\fi
\expandafter\ifx\csname urlprefix\endcsname\relax\def\urlprefix{URL }\fi
\providecommand{\bibinfo}[2]{#2}
\providecommand{\eprint}[2][]{\url{#2}}

\bibitem[{\citenamefont{M\"uck et~al.}(1998)\citenamefont{M\"uck, Andre,
  Clarke, Gail, and Heiden}}]{mueck1998}
\bibinfo{author}{\bibfnamefont{M.}~\bibnamefont{M\"uck}},
  \bibinfo{author}{\bibfnamefont{M.-O.} \bibnamefont{Andre}},
  \bibinfo{author}{\bibfnamefont{J.}~\bibnamefont{Clarke}},
  \bibinfo{author}{\bibfnamefont{J.}~\bibnamefont{Gail}}, \bibnamefont{and}
  \bibinfo{author}{\bibfnamefont{C.}~\bibnamefont{Heiden}},
  \bibinfo{journal}{Appl. Phys. Lett.} \textbf{\bibinfo{volume}{72}},
  \bibinfo{pages}{2885} (\bibinfo{year}{1998}).

\bibitem[{\citenamefont{M\"uck and McDermott}(2010)}]{mueck2010}
\bibinfo{author}{\bibfnamefont{M.}~\bibnamefont{M\"uck}} \bibnamefont{and}
  \bibinfo{author}{\bibfnamefont{R.}~\bibnamefont{McDermott}},
  \bibinfo{journal}{Supercon. Sci. Technol.} \textbf{\bibinfo{volume}{23}},
  \bibinfo{pages}{093001} (\bibinfo{year}{2010}).

\bibitem[{\citenamefont{M\"uck and Clarke}(2001)}]{mueck2001}
\bibinfo{author}{\bibfnamefont{M.}~\bibnamefont{M\"uck}} \bibnamefont{and}
  \bibinfo{author}{\bibfnamefont{J.}~\bibnamefont{Clarke}},
  \bibinfo{journal}{Appl. Phys. Lett} \textbf{\bibinfo{volume}{78}},
  \bibinfo{pages}{3666} (\bibinfo{year}{2001}).

\bibitem[{\citenamefont{M\"uck et~al.}(2003)\citenamefont{M\"uck, Welzel, and
  Clarke}}]{mueck2003}
\bibinfo{author}{\bibfnamefont{M.}~\bibnamefont{M\"uck}},
  \bibinfo{author}{\bibfnamefont{C.}~\bibnamefont{Welzel}}, \bibnamefont{and}
  \bibinfo{author}{\bibfnamefont{J.}~\bibnamefont{Clarke}},
  \bibinfo{journal}{Appl. Phys. Lett} \textbf{\bibinfo{volume}{82}},
  \bibinfo{pages}{3266} (\bibinfo{year}{2003}).

\bibitem[{\citenamefont{Clarke and Wilhelm}(2008)}]{clarke2008}
\bibinfo{author}{\bibfnamefont{J.}~\bibnamefont{Clarke}} \bibnamefont{and}
  \bibinfo{author}{\bibfnamefont{F.~K.} \bibnamefont{Wilhelm}},
  \bibinfo{journal}{Nature} \textbf{\bibinfo{volume}{453}},
  \bibinfo{pages}{1031} (\bibinfo{year}{2008}).

\bibitem[{\citenamefont{Wallraff et~al.}(2004)\citenamefont{Wallraff, Schuster,
  Blais, Frunzio, Huang, Majer, Kumar, Girvin, and Schoelkopf}}]{wallraff04}
\bibinfo{author}{\bibfnamefont{A.}~\bibnamefont{Wallraff}},
  \bibinfo{author}{\bibfnamefont{D.}~\bibnamefont{Schuster}},
  \bibinfo{author}{\bibfnamefont{A.}~\bibnamefont{Blais}},
  \bibinfo{author}{\bibfnamefont{L.}~\bibnamefont{Frunzio}},
  \bibinfo{author}{\bibfnamefont{R.}~\bibnamefont{Huang}},
  \bibinfo{author}{\bibfnamefont{J.}~\bibnamefont{Majer}},
  \bibinfo{author}{\bibfnamefont{S.}~\bibnamefont{Kumar}},
  \bibinfo{author}{\bibfnamefont{S.}~\bibnamefont{Girvin}}, \bibnamefont{and}
  \bibinfo{author}{\bibfnamefont{R.}~\bibnamefont{Schoelkopf}},
  \bibinfo{journal}{Nature} \textbf{\bibinfo{volume}{431}},
  \bibinfo{pages}{162} (\bibinfo{year}{2004}).

\bibitem[{\citenamefont{Serban et~al.}(2008)\citenamefont{Serban, Plourde, and
  Wilhelm}}]{serban08}
\bibinfo{author}{\bibfnamefont{I.}~\bibnamefont{Serban}},
  \bibinfo{author}{\bibfnamefont{B.~L.~T.} \bibnamefont{Plourde}},
  \bibnamefont{and} \bibinfo{author}{\bibfnamefont{F.~K.}
  \bibnamefont{Wilhelm}}, \bibinfo{journal}{Phys. Rev. B}
  \textbf{\bibinfo{volume}{78}}, \bibinfo{pages}{054507}
  (\bibinfo{year}{2008}).

\bibitem[{\citenamefont{Spietz et~al.}(2008)\citenamefont{Spietz, Irwin, and
  Aumentado}}]{spietz2008}
\bibinfo{author}{\bibfnamefont{L.}~\bibnamefont{Spietz}},
  \bibinfo{author}{\bibfnamefont{K.}~\bibnamefont{Irwin}}, \bibnamefont{and}
  \bibinfo{author}{\bibfnamefont{J.}~\bibnamefont{Aumentado}},
  \bibinfo{journal}{Appl. Phys. Lett.} \textbf{\bibinfo{volume}{93}},
  \bibinfo{pages}{082506} (\bibinfo{year}{2008}).

\bibitem[{\citenamefont{Tesche and Clarke}(1977)}]{tesche77}
\bibinfo{author}{\bibfnamefont{C.~D.} \bibnamefont{Tesche}} \bibnamefont{and}
  \bibinfo{author}{\bibfnamefont{J.}~\bibnamefont{Clarke}},
  \bibinfo{journal}{J. Low Temp. Phys.} \textbf{\bibinfo{volume}{29}},
  \bibinfo{pages}{301} (\bibinfo{year}{1977}).

\bibitem[{\citenamefont{Ketchen and Jaycox}(1982)}]{ketchen1982}
\bibinfo{author}{\bibfnamefont{M.~B.} \bibnamefont{Ketchen}} \bibnamefont{and}
  \bibinfo{author}{\bibfnamefont{J.~M.} \bibnamefont{Jaycox}},
  \bibinfo{journal}{Appl. Phys. Lett.} \textbf{\bibinfo{volume}{40}},
  \bibinfo{pages}{736} (\bibinfo{year}{1982}).

\bibitem[{\citenamefont{Ketchen et~al.}(1992)\citenamefont{Ketchen, Stawiasz,
  Pearson, Brunner, Hu, Jaso, Manny, Parsons, and Stein}}]{ketchen92}
\bibinfo{author}{\bibfnamefont{M.~B.} \bibnamefont{Ketchen}},
  \bibinfo{author}{\bibfnamefont{K.~G.} \bibnamefont{Stawiasz}},
  \bibinfo{author}{\bibfnamefont{D.~J.} \bibnamefont{Pearson}},
  \bibinfo{author}{\bibfnamefont{T.~A.} \bibnamefont{Brunner}},
  \bibinfo{author}{\bibfnamefont{C.-K.} \bibnamefont{Hu}},
  \bibinfo{author}{\bibfnamefont{M.~A.} \bibnamefont{Jaso}},
  \bibinfo{author}{\bibfnamefont{M.~P.} \bibnamefont{Manny}},
  \bibinfo{author}{\bibfnamefont{A.~A.} \bibnamefont{Parsons}},
  \bibnamefont{and} \bibinfo{author}{\bibfnamefont{K.~J.} \bibnamefont{Stein}},
  \bibinfo{journal}{Appl. Phys. Lett.} \textbf{\bibinfo{volume}{61}},
  \bibinfo{pages}{336} (\bibinfo{year}{1992}).

\bibitem[{\citenamefont{Kinion and Clarke}(2008)}]{kinion08}
\bibinfo{author}{\bibfnamefont{D.}~\bibnamefont{Kinion}} \bibnamefont{and}
  \bibinfo{author}{\bibfnamefont{J.}~\bibnamefont{Clarke}},
  \bibinfo{journal}{Appl. Phys. Lett.} \textbf{\bibinfo{volume}{92}}
  (\bibinfo{year}{2008}).

\bibitem[{\citenamefont{Dolan}(1977)}]{dolan1977}
\bibinfo{author}{\bibfnamefont{G.~J.} \bibnamefont{Dolan}},
  \bibinfo{journal}{Appl. Phys. Lett.} \textbf{\bibinfo{volume}{31}},
  \bibinfo{pages}{337} (\bibinfo{year}{1977}).

\bibitem[{\citenamefont{Kinion and Clarke}(2010)}]{kinion10}
\bibinfo{author}{\bibfnamefont{D.}~\bibnamefont{Kinion}} \bibnamefont{and}
  \bibinfo{author}{\bibfnamefont{J.}~\bibnamefont{Clarke}},
  \bibinfo{journal}{Appl. Phys. Lett.} \textbf{\bibinfo{volume}{96}},
  \bibinfo{pages}{172501} (\bibinfo{year}{2010}).

\bibitem[{\citenamefont{Wellstood et~al.}(1989)\citenamefont{Wellstood, Urbina,
  and Clarke}}]{wellstood89}
\bibinfo{author}{\bibfnamefont{F.~C.} \bibnamefont{Wellstood}},
  \bibinfo{author}{\bibfnamefont{C.}~\bibnamefont{Urbina}}, \bibnamefont{and}
  \bibinfo{author}{\bibfnamefont{J.}~\bibnamefont{Clarke}},
  \bibinfo{journal}{Appl. Phys. Lett.} \textbf{\bibinfo{volume}{54}},
  \bibinfo{pages}{2599} (\bibinfo{year}{1989}).

\bibitem[{\citenamefont{Wellstood et~al.}(1994)\citenamefont{Wellstood, Urbina,
  and Clarke}}]{wellstood94}
\bibinfo{author}{\bibfnamefont{F.~C.} \bibnamefont{Wellstood}},
  \bibinfo{author}{\bibfnamefont{C.}~\bibnamefont{Urbina}}, \bibnamefont{and}
  \bibinfo{author}{\bibfnamefont{J.}~\bibnamefont{Clarke}},
  \bibinfo{journal}{Phys. Rev. B} \textbf{\bibinfo{volume}{49}},
  \bibinfo{pages}{5942} (\bibinfo{year}{1994}).

\end{thebibliography}

\end{document}